\documentclass[10pt,twoside]{revtex4}%{amsart}
\usepackage[latin1]{inputenc}
\usepackage{t1enc}
\usepackage{amssymb,amsbsy,amsmath,amsfonts,amssymb,amscd}
\usepackage{latexsym,exscale}%,epic,eepic,epsfig}
\usepackage[english,french]{babel}
\usepackage{subeqnarray}
\usepackage[dvips]{graphicx}
\input utile.def
\newtheorem{e-proposition}[theorem]{Proposition}

\newtheorem{e-definition}[theorem]{D\'efinition\rm}

\newcommand{\nbre}{\nonumber}
\newcommand{\dhdt}{{\partial h\over\partial t}}
\newcommand{\hx}{{\partial h\over\partial x}}
\newcommand{\dt}{{\partial \over\partial t}}
\newcommand{\dx}{{\partial \over\partial x}}
\newcommand{\dudr}{{\partial u\over\partial r}}
\newcommand{\dudx}{{\partial u\over\partial x}}
\newcommand{\dvdr}{{\partial v\over\partial r}}
\newcommand{\dvdx}{{\partial v\over\partial x}}
\newcommand{\dpdx}{{\partial p\over\partial x}}

\newcommand{\dvdy}{{\partial v\over\partial y}}
\newcommand{\dudy}{{\partial u\over\partial y}}
\newcommand{\dpdy}{{\partial p\over\partial y}}
\newcommand{\dudyy}{{\partial^2  u\over\partial y^2}}
\newcommand{\dvdyy}{{\partial^2  v\over\partial y^2}}
\newcommand{\dudxx}{{\partial^2  u\over\partial x^2}}
\newcommand{\dudt}{{\partial u\over\partial t}}
\newcommand{\dvdt}{{\partial v\over\partial t}}
\newcommand{\hxx}{{\partial^2 h\over\partial x^2}}
\newcommand{\hxxx}{{\partial^3 h\over\partial x^3}}
\newcommand{\qx}{{\partial q\over\partial x}}
\newcommand{\qt}{{\partial q\over\partial t}}

\begin{document}

\title{Contribution to the modeling of a liquid film flowing down \\ inside
a vertical circular tube.\\
\vspace{1cm}
 Contribution à la modélisation d'un
film annulaire tombant.\\}
%Contribution à la modélisation d'un film liquide ruisselant à
%l'intérieur d'une conduite cylindrique verticale de section
%circulaire. }
%

\author{\textbf{
Samuel Ndoumbe~$^{\text{a,b}}$,\ \ François
Lusseyran~$^{\text{b}}$,\ \ Boujema Izrar~$^{\text{c}}$}}
\vspace{1cm}

\address{
\begin{itemize}\labelsep=2mm\leftskip=-5mm
\item[$^{\text{a}}$] Laboratoire d'Energétique de Mécanique
Théorique et Appliquée (LEMTA), CNRS-UMR 7563, 2, avenue de la
Forêt de Haye, BP 160, 54504 Vandoeuvre cedex, France.
 Courriel: ndoumbe@limsi.fr
\item[$^{\text{b}}$] Laboratoire d'Informatique pour la Mécanique
et les Sciences  de l'Ingénieur (LIMSI), CNRS-UPR 3251, Université
 Paris-Sud, 91403 ORSAY cedex, France.
Courriel: lussey@limsi.fr \item[$^{\text{c}}$] Laboratoire
d'Aérothermique, CNRS-UPR 9020, 3, avenue de la recherche
 Scientifique 45076  Orleans la Source, France.
Courriel: izrar@cnrs-orléans.fr.
\end{itemize}
Tel.: 33(0)1 69 85 81 72 Fax.: 33(0)1 69 85 80 88.}

%\address{
%$^{\text{a}}$] Laboratoire d'Energétique de Mécanique Théorique et
%Appliquée (LEMTA), CNRS-UMR 7563, 2, avenue de la Forêt de Haye,
%BP 160, 54504 Vandoeuvre cedex, France.
% Courriel: ndoumbe@limsi.fr\\
%$^{\text{b}}$] Laboratoire d'Informatique pour la Mécanique et les
%Sciences  de l'Ingénieur (LIMSI), CNRS-UPR 3251, Université
% Paris-Sud, 91403 ORSAY cedex, France.
%Courriel: lussey@limsi.fr \\
%$^{\text{c}}$] Laboratoire
%d'Aérothermique, CNRS-UPR 9020, 3, avenue de la recherche
% Scientifique 45076  Orleans la Source, France.
%Courriel: izrar@cnrs-orléans.fr. Tel.: 33(0)1 69 85 81 72 Fax.:
%33(0)1 69 85 80 88.}

\keywords{Modeling, annular film, two phase flow.}

%%%%%%%%%%%%%%%%%%
\begin{abstract}{%

This note focuses on the development of a $2D$ model of a thin
liquid film flowing down inside a vertical pipe. This model is
based on the large wavelength assumption and valid for high
Reynolds and  Weber numbers.\\

Un modèle $2D$ décrivant le ruissellement d'un film liquide mince
à l'intérieur d'une conduite cylindrique maintenue verticale est
développé. Il est basé sur l'hypothèse de grande longueur d'onde
et valable pour des grands nombres de Reynolds et de Weber. }
\end{abstract}

\maketitle

%\thispagestyle{empty}
%%%%%%%%%%%%%%%%%%%%%%%%%%%%%%%%%%%%%%%%%%%%%%%%%%%%%%%%%%%%
%%%  R\'esum\'e  %%%
%%%%%%%%%%%%%%%%%%%%
%\begin{Resume}{%
%Un modèle $2D$ décrivant le ruissellement d'un film liquide mince
%à l'intérieur d'une conduite cylindrique maintenue verticale est
%développé. Il est basé sur l'hypothèse de grande longueur d'onde
%et valable pour des grands nombres de Reynolds et de Weber.
% }\end{Resume}

%%%%%%%%%%%%%%%%%%%%%%%%%%%%%%%%%%%%%%%%%%%%%%%%%%%%%%%%%%%%
%%%  Abridged English version  %%%
%%%%%%%%%%%%%%%%%%%%%%%%%%%%%%%%%%
%\AEv
 \section{Introduction}
Various models relate to falling films on vertical or inclined
plate and small Reynolds number whereas most experimental
observations  at high Reynolds number are made inside cylindrical
tubes. The aim of this note is to present a $2D$ model which
accounts for small and medium cylinder curvature at high Reynolds
number, in order to make a more realistic comparison with the
experimental data \cite{Ndoumbe01}.
\section{Main equations}
 We consider a viscous fluid film flowing down a vertical
infinite pipe of  radius $R_c$ having a common interface with a
quiescent gas phase. System (\ref{ns1}-\ref{cl1}) defines the full
equations of motion for the annular falling film. Two small
parameters are defined: $\epsilon$ which is the ratio between the
film thickness $h_0$ and the characteristic length and
$\epsilon_r$ for the reduced curvature of the cylinder. In order
to reduce these model equations, we use a boundary-layer theory.
To satisfy the assumptions of large Reynolds numbers
($\displaystyle \epsilon\,Re=O(1)$ ) and $\displaystyle
\epsilon^2\,W=O(1)$, equations
(\ref{eqreduite}-\ref{sautpression}) are simplified by dropping
the terms of order $\epsilon/Re$ or higher. We assume that the
instantaneous streamwise velocity $u$  is described by a
self-similar velocity profile (\ref{uvitesse}). These reduced
equations are integrated through the local and instantaneous
thickness $h$ of the liquid film. The final model is a system of
two coupled equations (\ref{model2d}a, \ref{model2d}b) involving
$h$ and  the local flow rate $q$, as the only dependent variables.
We give an approximated model at order $\epsilon_r^2$.
\section{Discussion}
If one follows the Shkadov procedure \cite{Shkadov67} the
smallness parameter $\epsilon$ may be effectively included in the
dimensionless numbers of the flow, avoiding to specify a
characteristic length $l_c$. The fact remains that, in the
cylindrical geometry the parameter associated with the curvature
cannot be eliminated in the same way and the specification of
$l_c$ is mandatory. In this case, the characteristic length may be
defined as the length of the neutral mode \cite{Trifonov91} or the
length of the most energetic wave \cite{Ndoumbe01a}. The
cylindrical model developed in this study could be improved by
writing the streamwise velocity field as polynomials functions
\cite{Ruyer00}. The dimensionless equation are found  by using
time and space scales based only on the physical properties of
liquid which are the kinematic viscosity and the gravity
acceleration. This may contribute to give more information of the
nonlinear terms included in the functions $\Phi_1$ and $\Phi_2$.

\par\medskip\centerline{\rule{2cm}{0.2mm}}\medskip
\setcounter{section}{0}
%\selectlanguage{francais}
%%%%%%%%%%%%%%%%%%%%%%%%%%%%%%%%%%%%%%%%%%%%%%%%%%%%%%%%%%%%
%%%  Texte principal (en Francais)  %%%
%%%%%%%%%%%%%%%%%%%%%%%%%%%%%%%%%%%%%%%
\section{Introduction}
\indent Les divers développements théoriques réalisés sur le film
tombant se divisent en deux groupes. Le premier prend en compte
les deux phases de l'écoulement à travers des équations
d'évolution décrivant les variables de chaque phase. Le second
groupe, dans lequel nous nous plaçons, considère que la phase
gazeuse est inactive. Cependant, les différents modèles
\cite{Ruyer00} existant dans la littérature ne concernent que le
film ruisselant sur une plaque plane verticale ou inclinée alors
que la majeure partie des observations expérimentales
\cite{Karapantsios95}, compatibles avec des situations
industrielles, sont faites à l'intérieur des tubes cylindriques de
diamètre variant de $0.5$ à $10\,\rm{cm}$. Pourtant, on sait par
expérience que le cylindre est plus stable à haut débit que le cas
plan ce qui n'est pas explicable par la seule périodicité
azimutale. L'objet de cette note est de présenter un modèle $2D$
qui tienne compte de la courbure du cylindre et les grands nombres
de Reynolds afin de permettre une future comparaison plus réaliste
avec les données expérimentales \cite{Ndoumbe01}.

\section{Présentation du problème}
Nous considérons le ruissellement d'un film liquide mince le long
de la paroi interne d'une conduite cylindrique verticale (figure
\ref{cylindre}) de section circulaire et de rayon $R_c$ sous
l'effet de la gravité. Le liquide est newtonien incompressible  et
isotherme. Les points de départ de la recherche des équations
d'évolution du film sont les équations de Navier-Stokes:

\begin{figure}[h]
\begin{center}
  \includegraphics[scale=0.3]{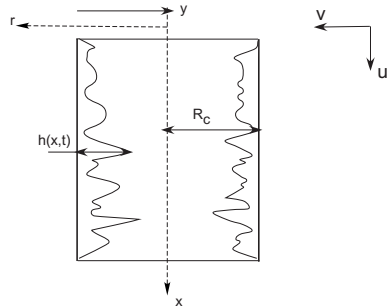}
 \caption{Représentation de la géométrie de l'écoulement.
 %\sl Sketch of the configuration and coordinate system.
 } \label{cylindre}
 \end{center}
\end{figure}

\begin{subeqnarray}\label{ns1}
&\displaystyle \stackrel\rightarrow\nabla\cdot\stackrel\rightarrow
 u=0 ,\\ %\quad \mbox{et}\quad
&\displaystyle \rho\left(\frac{\partial\stackrel\rightarrow
u}{\partial t}+ \stackrel\rightarrow u\stackrel
=\nabla\cdot\stackrel\rightarrow u\right)=\rho\stackrel\rightarrow
g -\stackrel\rightarrow\nabla p+\mu\triangle\stackrel\rightarrow
u,
\end{subeqnarray}
où $\rho$ est la densité, $\mu$ la viscosité dynamique, $p$ la
pression et $\stackrel\rightarrow u=(u,v)$ la vitesse. A ces
équations  sont jointes les conditions aux limites:\\
- de non glissement à la paroi:
\begin{equation}\label{eqreduite1}
\displaystyle u=v=0,
\end{equation}
- cinématique à l'interface:
\begin{equation}\label{condcinematique}
\displaystyle \dhdt+u\hx+v=0,
\end{equation}
- de saut dynamique à la surface libre:
\begin{subeqnarray}\label{cl1}
\displaystyle p-p_g=&\displaystyle -\frac{\sigma}{\displaystyle
\left[1+(\hx)^2\right]^{3/2}}\left[\hxx+\frac{1}{R_c-h}(1+(\hx)^2)\right]\\
\nbre &\displaystyle +\frac{2\mu}{\displaystyle
1+\hx^2}\left[\dvdr+\hx\left(\dudr+\dvdx\right)+\dudx(\hx)^2\right]\\
&\displaystyle
2\hx\left(\dudx-\dvdr\right)+\dudx\left(1-\hx^2\right)\left(\dudr+\dvdx\right)=0
\end{subeqnarray}
où  $\sigma$ est le coefficient de tension superficielle et $p_g$
la pression du gaz supposée constante.

\indent La solution de base correspond à un film tombant avec une
épaisseur constante $h_0$ le long de toute la conduite. Cette
solution  est obtenue à partir des équations précédentes en
supposant que l'écoulement est permanent ($\displaystyle \dt=0$)
et établi suivant l'axe ($\displaystyle v=0,\, \dx=0$) de la
conduite, dans ce cas nous obtenons des équations réduites dont la
solution est la composante axiale de la vitesse
\begin{equation}\label{ubase}
\displaystyle u=\frac{g\,R_{c}^{2}}{4\nu}
\left[1-\left(\frac{r}{R_c}
\right)^2+2\left(1-\frac{h_{0}}{R_c}\right)^2
\log\left(\frac{r}{R_c}\right)\right],
\end{equation}
où $\nu$ est la viscosité cinématique. Le débit est
\begin{equation}\label{debitm}
q_{0}=\int_{R_c-h_{0}}^{R_c}2\,\pi r\,u\,dr
\end{equation}
correspondant à une vitesse débitante
\begin{equation}\label{vmoy}
\displaystyle u_{0}=\frac{q_{0}}{\displaystyle \pi
R_c^2\left(1-\left(1-\displaystyle
\frac{h_{0}}{R_c}\right)^2\right)}.
\end{equation}
L'épaisseur $h_{0}$ est solution de l'équation
%\begin{center}
\begin{equation}\label{hbase}
 \begin{array}{ll}
&\displaystyle 8\,\nu^2\,Re\,\left(R_c^2-\left(R_c-\displaystyle
h_0\right)^2\right)=g\,h_0\,R^{4}_{c}\times\\
&\displaystyle \hspace{1cm}\left[1-4
\left(1-\frac{h_0}{R_c}\right)^2+
3\left(1-\frac{h_0}{R_c}\right)^4-4\left(1-\frac{h_0}{R_c}
\right)^4\log\left(1-\frac{h_0}{R_c}\right)\right]
\end{array}
\end{equation}
%\end{center}
où $Re=u_{0}\,h_{0}/\nu$ est le nombre de Reynolds. L'écart entre
$h_0$ et l'épaisseur de Nüsselt $\displaystyle
\left(3\,\nu^2\,Re/g\right)^{1/3}$ est négligeable pour $R_c >
1.5\,\rm{cm}$.

Dans la suite, nous introduisons une nouvelle variable $y=R_c-r$,
l'expression (\ref{ubase}) de la vitesse devient
\begin{equation}\label{vplan}
\displaystyle u=\frac{g\,R^{2}_{c}}{2\nu} \left[\frac{y}{R_c}
-\frac{1}{2}\left(\frac{y}{R_c}
\right)^2+\left(1-\frac{h_0}{R_c}\right)^2
\log\left(1-\frac{y}{R_c}\right)\right],
\end{equation}
où $\displaystyle \frac{g\,R^{2}_{c}}{2\nu} \left(\frac{y}{R_c}
-\frac{1}{2}\left(\frac{y}{R_c} \right)^2\right)$ correspond au
profil de base d'un film liquide ruisselant sur un plan vertical
et $ \displaystyle \frac{g\,R^{2}_{c}}{2\nu}
\left(\left(1-\frac{h_0}{R_c}\right)^2
\log\left(1-\frac{y}{R_c}\right)\right)$ est le terme de
correction du profil plan lié à la courbure du cylindre. La paroi
du cylindre est identifiée par l'équation $\displaystyle y=0$ et
l'interface par la relation $\displaystyle y=h(x,t)$

\indent Pour tenir compte de l'hypothèse de grande longueur d'onde
\cite{Izrar93}, nous introduisons le rapport d'aspect
$\displaystyle \epsilon=h_0/l_c$ où $l_c$ est la longueur
caractéristique suivant l'axe des $x$. Nous négligeons les termes
d'ordre supérieur à $\epsilon^2$ pour aboutir aux équations
réduites de Navier-Stokes sous la forme adimensionnelle:
\begin{subeqnarray}\label{eqreduite}
&\displaystyle \dudx-\dvdy+\frac{v}{R-y}=0\\
 &\displaystyle -\frac{1}{Re}\left(-\frac{1}{R-y}\dudy
+\dudyy\right)-\frac{1}{Fr^2}+\left(\dudt+u\dudx-v\dudy+\dpdx
\right)\epsilon-\frac{\epsilon^2}{Re}\dudxx=0\\
&\displaystyle -\dpdy-\frac{\epsilon}{Re}\left(-\frac{1}{R-y}\dvdy
+\dvdyy-\frac{v}{\left(R-y\right)^2}\right)+\left(\dvdt+u\dvdx-v\dvdy
\right)\epsilon^2=0
\end{subeqnarray}
La condition d'adhérence et les conditions aux limites restent
inchangées alors que les conditions dynamiques donnent
\begin{subeqnarray}\label{sautpression}
&\displaystyle p= p_g-
 \frac{W}{R-h}+\frac{2\epsilon}{Re}\left(-\dvdy+\hx\dudy\right)-
\,\epsilon^2\,W\left(\displaystyle \hxx
 -\frac{1}{2(R-h)}\left(\hx\right)^2\right)\\
&\displaystyle \frac{1}{Re}\dudy+\frac{\epsilon^2}{Re}\left[
 -\dvdx+2\hx\left(\dvdy+\dudx\right)\right]=0.
\end{subeqnarray}
où $\displaystyle W=\sigma/\left(\rho\,h_0\,u_0^{2}\right)$ est le
nombre de Weber, $R=R_c/h_0$ le rayon adimensionnel et
$Fr=u_0/\sqrt{gh_0}$ le nombre de Froude qui prend la valeur
$\sqrt{Re/3}$ en géométrie plane verticale. Dans les équations
(\ref{eqreduite} - \ref{sautpression}), les différentes variables
ont été réduites de la manière suivante:
\begin{equation}\label{adimen}
x\rightarrow x/l_c,\; r\rightarrow R- y/h_0,\;t\rightarrow t/
(l_c/u_0),\; u\rightarrow u/u_0,\; v\rightarrow \epsilon v/u_0,\;
p\rightarrow p/(\rho u_{0}^2).
\end{equation}
Il est à noter que $\displaystyle \dt \sim \dx \sim \epsilon $.
L'hypothèse de grande longueur d'onde impose $\epsilon$ petit
devant l'unité.  En faisant l'hypothèse que $\epsilon Re= O(1)$ et
$\epsilon^2 W= O(1)$ , nous pouvons négliger dans les équations
(\ref{eqreduite}b, \ref{eqreduite}c, \ref{sautpression}a,
\ref{sautpression}b) le terme en $\epsilon/Re$, $\epsilon^2/Re$ et
$\epsilon^2$, pour obtenir:
\begin{subeqnarray}\label{eqreduite1}
&\displaystyle \dudx-\dvdy+\frac{v}{R-y}=0\\
 &\displaystyle -\frac{1}{Re}\left(-\frac{1}{r}\dudy
+\dudyy\right)-\frac{1}{Fr^2}+\left(\dudt+u\dudx-v\dudy+\dpdx
\right)\epsilon=0\\
&\displaystyle \dpdy=0\\
 &\displaystyle p|_{y=h}= p_g-
 \frac{W}{R-h}-
\,\epsilon^2\,W\left(\displaystyle \hxx
 -\frac{1}{2(R-h)}\left(\hx\right)^2\right)\\
&\displaystyle \dudy|_{y=h}=0.
\end{subeqnarray}

\section{Équations d'évolution du film tombant}
Il apparaît donc naturel de ramener toutes les équations  réduites
(\ref{eqreduite} - \ref{sautpression}) à l'interface par
intégration sur l'épaisseur. Cette technique est d'usage courant
dans les problèmes où l'inconnue principale est localisée sur une
partie de la frontière du domaine. Le relèvement des équations
(\ref{eqreduite}a, \ref{eqreduite}b) sur l'épaisseur se fait avec
l'hypothèse que le profil de vitesse est auto similaire
\begin{equation}\label{uvitesse}
\displaystyle u\left(x,y,t\right)=U\left(x,t\right)\,\frac{\varphi
\displaystyle
\left(y\right)}{{\varphi\left(h\left(x,t\right)\right)}}
\end{equation}
\noindent où $\displaystyle
\varphi(y)=y/R-1/2\left(y/R\right)^2+\left(1-h/R\right)^2\ln\left(1-{y}/{R}\right)$
et $U$ la vitesse locale à l'interface du film. On pose $\psi
(y)=\varphi \displaystyle
\left(y\right)/\varphi\left(h\left(x,t\right)\right)$. A partir de
la définition de $\displaystyle
q=\int_{0}^{h}\left(1-\frac{y}{R}\right) \,u\,dy$, le débit
instantané du film, le relèvement intégral consiste à écrire
(\ref{eqreduite1}a, \ref{eqreduite1}b) sous la forme
\begin{subeqnarray}\label{eqint}
\displaystyle
\dx\left(\int_{0}^{h}\left(R-y\right)\,u\,dy\right)+\left(R-h\right)\left(\hx u|_{y=h}+v|_{y=h}\right)=0\\
\displaystyle R\qt+\dx\left(\int_{0}^{h}\left(R-y\right)\,u^2
dy\right)-
\frac{R}{\epsilon\,Re}\dudy|_{y=0}+\left(\frac{h^2}{2}-h\,R\right)\,\left(\dx\left(\displaystyle
p\mid_{y=h}\right)-\frac{1}{\epsilon\,Fr^2}\right)=0
\end{subeqnarray}
En tenant compte des conditions aux limites
(\ref{eqreduite1}-\ref{sautpression}), on aboutit donc à un
système de deux équations dont les inconnues sont $h$ l'épaisseur
instantanée et $q$:
\begin{subeqnarray}\label{model2d}
&\displaystyle \left(1-\frac{h}{R}\right)\dhdt+\displaystyle \qx=0\\
&\displaystyle
\qt+\dx\left(q^2\,\Phi_1\left(h\right)\right)-\frac{q\,\Phi_2\left(h\right)}{\epsilon\,Re}+
\left(h-\frac{h^2}{2\,R}\right)\,\left(\dx\left(\displaystyle
p\mid_{y=h}\right)-\frac{1}{\epsilon\,Fr^2}\right)=0
\end{subeqnarray}
où $ \displaystyle \int_{0}^{h}\left(R-y\right)\, u^2 dy\,=\,
\displaystyle R q^2\Phi_1(h),\quad \displaystyle \dudy|_{y=0}
\,=\,\displaystyle q \Phi_2(h)$. Les expressions des fonctions
$\Phi_1$ et $\Phi_2$ sont facilement déterminées par les relations

 $\displaystyle
\Phi_1\left(h\right)=R^2\left(\int_{0}^h
\psi^2\left(\zeta\right)\zeta d\zeta\right)/\left(\int_{0}^h
\psi\left(\zeta\right)\zeta d\zeta \right)^2$  et $ \displaystyle
\Phi_2\left(h\right)=R\, \psi\,'(0)/\left(\int_{0}^h
\psi\left(\zeta\right)\zeta d\zeta\right).$

\noindent En définissant un nouveau petit paramètre
$\epsilon_r=1/R$ qui mesure les effets de courbure, nous faisons
un développement en série des équations
(\ref{model2d}a-\ref{model2d}b)  et obtenons:
\begin{subeqnarray}\label{model2d1}
&\displaystyle \dhdt+\displaystyle\left(1+h\epsilon_r+h^2\epsilon_r^2\right) \qx=0\\
&\displaystyle \qt+\dx\left(A\left(\epsilon_r
h\right)\,q^2\right)-B\left(\epsilon_r
h\right)\frac{q}{\epsilon\,Re}- C\left(\epsilon_r
h\right)\frac{1}{\epsilon\,Fr^2}\\ \nbre &\displaystyle
+C\left(\epsilon_r h\right)\left[ \epsilon^2 W
\left(\hxxx-\hx\hxx\epsilon_r-\frac{1}{2}\hx^3-h\hx\hxx\epsilon_r^2
\right)-W\hx\epsilon_r^2\right]=0
\end{subeqnarray}
où $\displaystyle A\left(\epsilon_r
h\right)=\frac{6}{5h}+\frac{7}{10}\epsilon_r+\frac{139}{350}h\epsilon_r^2,
\quad \displaystyle  B\left(\epsilon_r
h\right)=-\frac{3}{h^2}-\frac{3}{2h}\epsilon_r-\frac{21}{20}\epsilon_r^2,\quad
\displaystyle C\left(\epsilon_r h\right)=2h-h^2\epsilon_r.$

\noindent On retrouve le cas plan en faisant tendre vers l'infini
le rayon de la conduite, ce qui redonne le modèle de Shkadov
correspondant aux équations (\ref{model2d1}a,\ref{model2d1}b) avec
$\epsilon_r=0$.

\section{Discussion et conclusion}
Ce modèle décrivant le ruissellement d'un film liquide dans une
conduite cylindrique verticale  permet d'explorer la dynamique
linéaire du film dans des gammes de nombres de Reynolds
correspondant à un régime d'écoulement interne laminaire ($\leq
700$).

\indent Si on suit la démarche de Skhadov \cite{Shkadov67} on peut
certes inclure le paramètre de petitesse dans les nombres
adimensionnels de l'écoulement et ainsi éviter de spécifier une
échelle de longueur axiale. Ce n'est plus le cas en géométrie
cylindrique où les termes de courbure imposent d'expliciter $l_c$.
Dans ce cas, on peut définir la longueur caractéristique comme
étant la longueur du mode neutre \cite{Trifonov91} ou la longueur
d'onde de l'onde la plus énergétique \cite{Ndoumbe01a}. Une autre
démarche consiste d'une part à prendre comme échelle de longueur
caractéristique $l_c$ construite à partir de deux grandeurs
physiques telles que $\nu$ et $g$ \cite{Ruyer00} et d'autre part à
introduire un nouveau petit paramètre lié à la courbure du
cylindre dont l'expression sera $\displaystyle l_c/R_c$. Cette
démarche permettrait d'obtenir un modèle dont la construction se
ferait à partir du profil de vitesse semi-parabolique dont la
forme polynômiale s'adapte bien aux développements  déjà initiés
par \cite{Ruyer00} et d'avoir accès aux termes en $\epsilon^2$ et
ceux propres à la configuration cylindrique et qui sont englobés
dans $\Phi_1$ et $\Phi_2$. En pratique le film plan est réputé
moins résistant aux perturbations extérieures (vibrations, action
de l'écoulement gazeux,...) que le film annulaire. Pourtant, du
point de vue théorique, en considérant la très faible influence du
rayon de courbure sur l'épaisseur moyenne du film, il est
généralement considéré que son rôle est négligeable. Il serait
donc intéressant, en partant d'une analyse linéaire des
instabilités axiales, de quantifier l'influence du rayon sur les
taux de croissance.

%%%%%%%%%%%%%%%%%%%%%%%%%%%%%%%%%%%%%%%%%%%%%%%%%%%%%%%%%%%%
%%%  Remerciements  %%%
%%%%%%%%%%%%%%%%%%%%%%%
%\Remerciements{Les remerciements.}
%%%%%%%%%%%%%%%%%%%%%%%%%%%%%%%%%%%%%%%%%%%%%%%%%%%%%%%%%%%%
%%%  Bibliographie %%%
%%%%%%%%%%%%%%%%%%%%%%

\end{document}